# Evaluating the Assessment of Software Fault-Freeness

John Rushby
Computer Science Laboratory
SRI International, Menlo Park CA USA

Bev Littlewood and Lorenzo Strigini
Centre for Software Reliability
City University London, UK

## Abstract

We propose to validate experimentally a theory of software certification that proceeds from assessment of confidence in fault-freeness (due to standards) to conservative prediction of failure-free operation.

## 1. Introduction

The quality required of safety-critical software is such that faults are not expected to be revealed by post-construction assurance processes, nor are failures expected in operation. Hence, we cannot expect to evaluate standards for software assurance by measuring reductions in faults or failures. Before we can frame testable hypotheses about standards, we therefore need to posit a larger hypothesis that evidence for absence of faults provides the quantifiable basis for certification.

## 2. A Theory of Certification

The world is an uncertain place, so safety require-ments are expressed probabilistically: it must be very unlikely that the system will lead to harm—and the more serious the harm, the less likely it must be. These likelihoods may be expressed in terms of probability of failure on demand (pfd), as rates of failure, or in terms of total exposure. We will use the latter, which, for the case of commercial aircraft, is expressed by requirements such as "no catastrophic failure condition shall be expected to occur in the entire lifetime of all aircraft of one type."

It might then seem that the purpose of standards and assurance for safety-critical software should be to deliver direct guarantees that this or similar goals are satisfied. However, a little reflection shows that software assurance cannot, and does not, accom-plish this direct demonstration. Failure is a dynamic concept—that is, it concerns the behavior of software in execution—and probabilistic claims about failure concern repeated executions. But software assurance is largely comprised of static forms of analysis: for exam-ple, examination of requirements, specifications, and code, and traceability among these. These analyses can discover faults and thereby improve software quality and reliability butit is not expected that any faults will be detected during system assurance: if they were, it would cast serious doubt on the development process. But if the value of assurance is not in revealing faults, how does it connect to failures and their probabilities? The answer is that it does so indirectly: assurance gives us confidence that faults are rare and from this we infer that the probability of failure will be low also. But how can we quantify this purported relationship?

We do more assurance for software with more stringent failure requirements (e.g., DO178C specifies 71 assurance "objectives" for Level A software, vs. 69 for Level B, 62 for Level C, and 26 for Level D) and the purpose of doing more assurance must be to make us more confident—but confident about what? We can imagine two answers: more assurance makes us confident in fewer faults and thereby a lower probability of failure (i.e., more assurance changes what we are confident about); or, it makes us more confident in a given rarity of faults and an associated probability of failure (i.e., it increases our confidence in a single, fixed, property). The first might seem more intuitive but it is a special case of the second that delivers a tractable theory.

The special case is software that is believed to be entirely free of faults (of kinds that could lead to failures of the severity under consideration). That is, more assurance makes us more confident that the soft-ware is fault-free and our degree of confidence can be expressed as a probability, namely $P(\text{s/w fault-free})$. By the formula for total probability

$$P(\text{s/w fails [on a randomly selected demand]}) \quad (1)$$
$$= P(\text{s/w fails} \mid \text{s/w fault-free}) \cdot P(\text{s/w fault-free})$$
$$+ P(\text{s/w fails} \mid \text{s/w faulty}) \cdot P(\text{s/w faulty}).$$

The first term in this sum is zero, because the software does not fail if it is fault-free (which is why the theory needs this property). Hence, if we define $p_{nf}$ as the probability the software is fault-free (or nonfaulty, so that $P(\text{s/w faulty}) = 1 - p_{nf}$), and $p_{F|f}$ as the probability that it Fails, if faulty, then $pfd = p_{F|f}(1 - p_{nf})$:

More importantly, $p_{srv}(n)$, the probability of surviving $n$ independent demands (e.g., flights) without failure is given by

$$p_{srv}(n) = p_{nf} + (1 - p_{nf})(1 - p_{F|f})^n \quad (2)$$

A suitably large $n$ can represent "the entire lifetime of all aircraft of one type." The notable feature of (2) is that the first term establishes a lower bound for $p_{srv}(n)$ that is independent of $n$. Thus, if software assurance gives us the confidence to assess, say, $p_{nf} > 0.99$ (or whatever threshold "not expected to occur" means) then it looks like we have sufficient evidence to certify the aircraft as safe (with respect to software aspects).

But certifiers (and the public) will also want guarantees in case the software does have faults. Thus, we need confidence that the second term in (2), which decays exponentially, will be well above zero. This confidence could come from prior failure-free operation (e.g., flight tests). Calculating the overall $p_{srv}(n)$ can then be posed as a problem in Bayesian inference: we have assessed a value for $p_{nf}$, have observed some number $r$ of failure-free demands, and want to predict the probability of seeing $n - r$ future failure-free demands. To do this, we need a prior distribution for $p_{F|f}$, which may be difficult to obtain, and difficult to justify for certification. However, there is a distribution that delivers provably worst-case predictions (specifically, one in which $p_{F|f}$ is concentrated in a probability mass at some $q_n \in (0;1]$) [1] so we can make predictions that are guaranteed to be conservative given only $p_{nf}$, $r$, and $n$. For values of $p_{nf}$ above $0.9$, we find that $p_{srv}(n)$ is well above the floor given by $p_{nf}$, provided $r > 10^n$.

If we regard a complete flight as a demand, then $n$ might be as large as $10^8-10^9$ (e.g., the Airbus A320 series have already performed over 62 million flights), but flight tests prior to certification might provide only $r = 10^3$, so it looks as if these are insufficient for certification by the criterion above. However, it can be argued that when an aircraft type is certified, we do not require (and in fact cannot feasibly obtain) sufficient evidence to predict failure-free operation over the entire lifetime of the type; instead, we initially require sufficient confidence for only, say, the first six months of operation and the small number of aircraft that will be deployed in that period. This will be a much smaller value of $n$, and our $p_{nf}$ (from assurance) and our $r$ (from flight tests) will be sufficient for confidence in its failure-free operation. Then we will need confidence in the next six months of operation, with a larger fleet, (i.e., a larger $n$) but now have the experience of the prior six months failure-free operation (i.e., a larger $r$) and in this way we can "bootstrap" our way forward.

## 3. Suggested Experiments

We contend that the account of the previous section provides the first scientific explanation for the way software certification in some industries (particularly commercial aviation) actually works: that is, it provides a model for beliefs and their quantification that explains and is consistent with current practice.

The hypotheses we propose to examine are (a) do the numbers work, and (b) do certifiers find the account plausible—to the extent that it can be used as a foundation on which to improve future practice?

For the first, we need to ask what values of $p_{nf}$ can reasonably be assessed for the various DO-178C Software Levels. Two approaches are: (i) ask certifiers what $p_{nf}$, cast in a frequentist interpretation, they might assess for each group of objectives: e.g., "given 100 software systems assessed to have accomplished all 7 objectives of DO-178C Section 6.3.2, how many of those systems do you believe might ever suffer a software failure due to flawed low level requirements?"; (ii) consider how many such systems have been in use and never exhibited such failures. Both approaches have (different) weaknesses, but allow construction of a first-cut plausible argument, e.g., using Bayesian Belief Nets and suitable conservative simplifications, to yield assessment of $p_{nf}$ for the whole of DO-178C.

We will also build models for the growth in fleet size over time and the "bootstrapping" of confidence in future failure-free operation (i.e., as $r$ and $n$ increase) for representative values of $p_{nf}$.

For hypothesis (b), we will present the model, peer-reviewed and populated with "plausible" data from (a), to certifiers. If they accept the structure of the argument then we may proceed to investigate methods by which its parameters, the $p_{nf}$ values, can be based rigorously on evidence. Possible approaches include analysis of the assurance case underlying the objectives of DO-178C and of means for accomplishing them. Further explorations could include modified objectives and alternative means (e.g., software monitors) to support assessment of high $p_{nf}$, enhancing $p_{srv}(n)$.